\documentclass[journal,twoside, print]{ieeecolor}
\usepackage{tmi}
\usepackage{cite}
\usepackage{amsmath,amssymb,amsfonts}
\usepackage{algorithmic}
\usepackage{graphicx}
\usepackage{textcomp}

\usepackage{booktabs}
\def\BibTeX{{\rm B\kern-.05em{\sc i\kern-.025em b}\kern-.08em
    T\kern-.1667em\lower.7ex\hbox{E}\kern-.125emX}}
\begin{document}
\title{Unsupervised PET Reconstruction from a Bayesian Perspective}
\author{Chenyu Shen, Wenjun Xia, Hongwei Ye, Mingzheng Hou, Hu Chen, Yan Liu, Jiliu Zhou, \IEEEmembership{Senior Member, IEEE}, and Yi Zhang, \IEEEmembership{Senior Member, IEEE}
\thanks{C. Shen, W. Xia, M. Hou, H. Chen, J. Zhou and Y. Zhang are with the
College of Computer Science, Sichuan University, Chengdu 610065,
China (e-mail: fx1024cy@gmail.com, xwj90620@gmail.com, houmingzheng@scu.edu.cn, huchen@scu.edu.cn, zhoujl@scu.edu.cn, yzhang@scu.edu.cn).}
\thanks{H. Ye is with the MinFound Medical Systems Co., Ltd, Hangzhou 310018, China (e-mail:  Hongwei.Ye@minfound.com).}
\thanks{Y. Liu is with the College of Electrucal Engineering, Sichuan University,
Chengdu 610065, China (e-mail: liuyan77@scu.edu.cn).}}

\maketitle

\begin{abstract}
    Positron emission tomography (PET) reconstruction has become an ill-posed inverse problem due to low-count projection data, and a robust algorithm is urgently required to improve imaging quality. Recently, the deep image prior (DIP) has drawn much attention and has been successfully applied in several image restoration tasks, such as denoising and inpainting, since it does not need any labels (reference image). However, overfitting is a vital defect of this framework. Hence, many methods have been proposed to mitigate this problem, and DeepRED is a typical representation that combines DIP and regularization by denoising (RED). In this article, we leverage DeepRED from a Bayesian perspective to reconstruct PET images from a single corrupted sinogram without any supervised or auxiliary information. In contrast to the conventional denoisers customarily used in RED, a DnCNN-like denoiser, which can add an adaptive constraint to DIP and facilitate the computation of derivation, is employed. Moreover, to further enhance the regularization, Gaussian noise is injected into the gradient updates, deriving a Markov chain Monte Carlo (MCMC) sampler. Experimental studies on brain and whole-body datasets demonstrate that our proposed method can achieve better performance in terms of qualitative and quantitative results compared to several classic and state-of-the-art methods.
\end{abstract}


\section{Introduction}
\label{sec:introduction}
\IEEEPARstart{P}{ositron} emission tomography (PET) is an important imaging modality in the field of nuclear medicine that can reflect the molecular-level activities of various organs and tissues in vivo by injecting specific radioactive tracers. Since the amount of injected radiotracer in current protocols will increase the potential risks of radiation exposure for the human body, reducing the dose is of clinical significance. However, simply reducing the injected dose can result in low-count PET projection data (sinogram), which leads to undesired noise and artifacts in the reconstructed images, further negatively affecting the subsequent clinical diagnosis. How to reconstruct high-quality images has always been a popular topic, and extensive efforts have been devoted over the past decades. These algorithms can be roughly classified into three groups: 1) analytical method; 2) image postprocessing; and 3) model-based iterative reconstruction (MBIR).

The most famous analytical method is filtered back-projection (FBP) \cite{a4}, which applies a convolution filter to remove blurring and relies on the central slice theorem. However, FBP is famously susceptible to measurement noise.

A simple solution for this problem is the postprocessing method. This kind of method directly processes the reconstructed PET images with image restoration algorithms, such as nonlocal mean (NLM) \cite{a17}, BM3D \cite{a18} and multiscale transform \cite{a68}. Although these methods have no requirement to access the raw data and have a relatively fast processing speed, the latent information in the sinogram is ignored, and the results have potential risks of being oversmoothed.

To tackle the disadvantages of analytical and postprocessing methods, MBIR methods have become mainstream in the past two decades. Maximum-likelihood expectation maximization (ML-EM) \cite{a5} is a classic iterative method that maximizes the log-likelihood of projection data. However, such an iterative reconstruction method without regularization formed by specific prior knowledge about images or sinograms may increase the noise after a certain number of iterations. As a result, integrating MBIR with specific regularization is desirable to suppress noise efficiently. Numerous different types of regularization have been proposed. Some typical methods include total variation \cite{a6}--\cite{a9}, edge-preserving filtering \cite{a10, a11}, dictionary learning \cite{a12}, kernel methods \cite{a13} and anatomical priors \cite{a14}--\cite{a16}. Despite promising results achieved by these methods, heavy computational cost and laborious parameter adjustment severely hinder their clinical applications. Moreover, since the construct of the regularization term is based on specific prior knowledge, it is difficult to find a universal regularization term, which can be easily adapted to different cases.

\begin{figure*}[htb]

  \centering
  \centerline{\includegraphics[width=13cm]{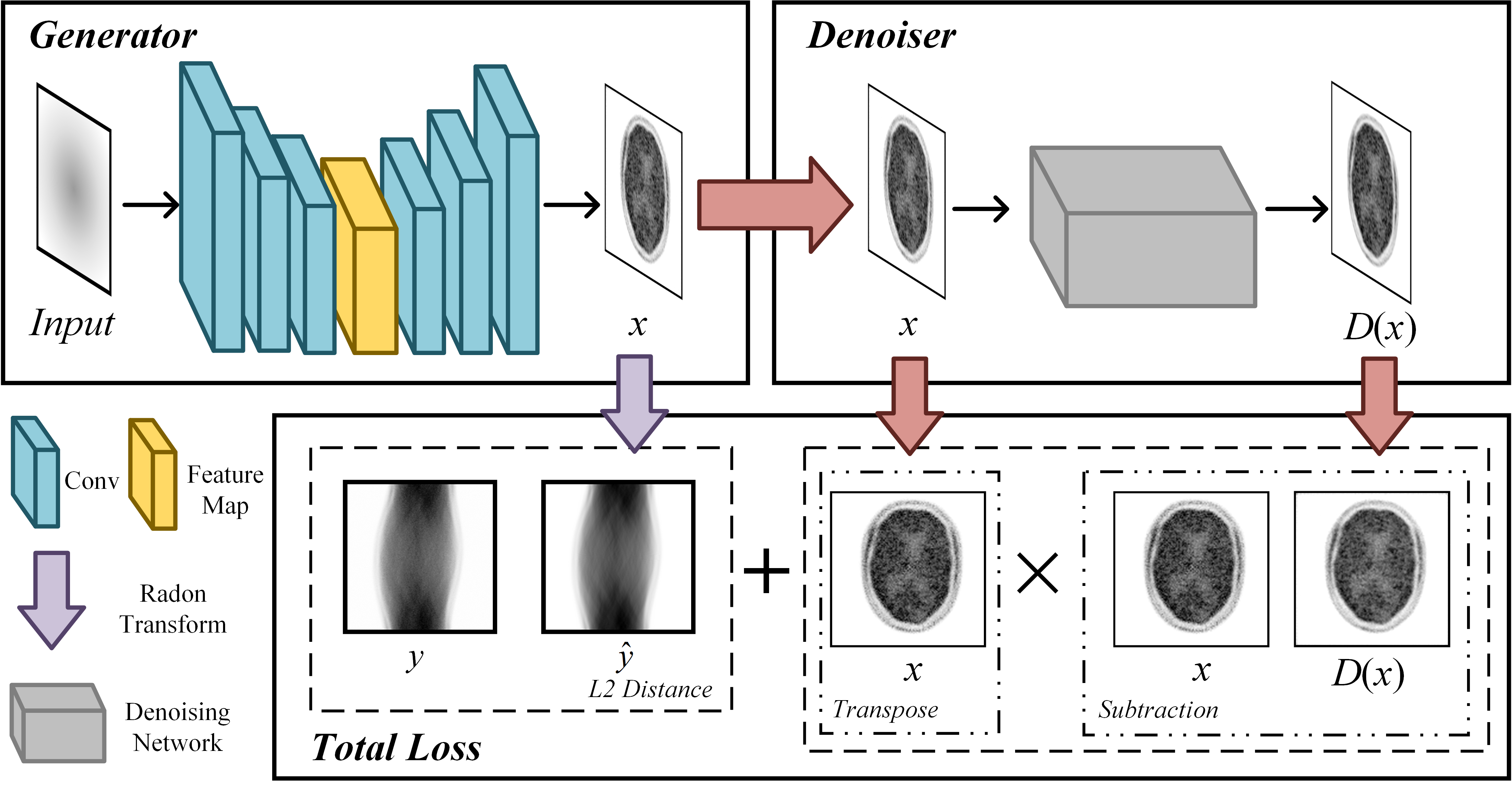}}
%  \vspace{2.0cm}
  \caption{Overview of our proposed model. The detailed generator and denoiser are shown in Fig. 2 and Fig. 3, respectively.}
  \label{fig:net}
  \end{figure*}

Over the past decade, deep neural networks have dramatically improved the state of the art in different fields of computer vision, e.g., object detection \cite{a19}, image recognition \cite{a20}, image segmentation \cite{a21}, and even medical imaging. For instance, convolutional neural networks (CNNs) and generative adversarial networks (GANs) have been applied in CT and MRI reconstruction \cite{a22, a23, a55, a56, a57, a58}, which have shown competitive performance to traditional methods. Recent studies on PET imaging reconstruction based on deep learning can be roughly divided into three groups. 1) Directly learn the mapping from the sinogram to the corresponding PET image. For example, DeepPET \cite{a24} was proposed to employ the sinogram as input and outputs PET images directly based on encoder-decoder network architecture. Furthermore, Hu \textit{et al.} \cite{a25} used an improved Wasserstein generative adversarial network aided by perceptual loss to reconstruct PET images from sinograms, which alleviated the problem of oversmoothness and detail loss that usually appeared in plain networks.
%  Kandarpa et al. \cite{a59} proposed a direct reconstruction framework composed of three steps: denoising, reconstruction and superresolution. The denoising and superresolution parts act as the data domain and image domain processing, respectively, while the reconstruction part acts as the data-to-image transform. 
 2) Learn the mapping from low-dose or low-count images to their corresponding full-dose or full-count images \cite{a26, a27, a62, a64}, which can be treated as image restoration or enhancement tasks. 
% Additionally, the estimate from low-dose PET to high-dose PET can be combined with other tasks, such as motion correction, to simultaneously denoise low-dose images and correct the motion \cite{a60}. 
3) Iterative reconstruction networks, which combine a neural network with MBIR by unrolling the iterative algorithm or substituting the regularization term with a learning-based prior. Typically, inspired by the kernel method, Gong \textit{et al.} \cite{a28} embedded a trained network into a constrained optimization problem and solved it using the alternating direction method of multipliers (ADMM). Mehranian \textit{et al.} \cite{a29} proposed a forward-backward splitting algorithm to integrate deep learning into the framework of MAP reconstruction, which is composed of regularization, expectation-maximum and weighted fusion. BCD-Net \cite{a65} is another iterative neural network constructed by unfolding a block coordinate descent MBIR algorithm, showing significant improvement in extreme low count PET reconstruction \cite{a66}. All these abovementioned methods are supervised and require large amounts of samples to train a proper network. However, it is not always feasible in clinical practice, especially for pilot clinical trials, to obtain such an amount of training samples, especially for paired low-dose or low-count data and their corresponding normal-dose or normal-count data.

To overcome this obstacle, recently, a trend of incorporating the well-known unsupervised learning model, called deep image prior (DIP) \cite{a30}, has come to the fore. DIP demonstrates that CNNs without pretraining can capture low-level statistics intrinsically. CNNs’ ability to solve inverse problems in the field of image processing, such as denoising and super resolution, is due to the network architecture itself instead of learning. Gong \textit{et al.} \cite{a31} integrated DIP into PET reconstruction while optimizing a Poisson distribution-based log-likelihood function. Yokota \textit{et al.} \cite{a32} incorporated DIP with non-negative matrix factorization for dynamic PET reconstruction. Under the framework of DIP, Cui \textit{et al.} employed the patient’s CT/MR images as the auxiliary input of conditional GAN to improve the performance of the original GAN \cite{a67}. Nonetheless, without early stopping or an appropriate stop criterion, DIP is apt to overfit. Designing a proper stop criterion is the key point to ease this problem. Mataev \textit{et al.} \cite{a33} proposed the DeepRED framework, merging the concept of regularization by denoising (RED) \cite{a34} into DIP by introducing an explicit prior. Cheng \textit{et al.} \cite{a35} provided a novel Bayesian view of DIP by conducting posterior sampling based on stochastic gradient Langevin dynamics (SGLD) \cite{a36}. DeepRED and SGLD were applied to PET reconstruction in \cite{a37} and \cite{a38}, respectively. However, DeepRED usually takes BM3D or NLM as its denoiser, which suffers from two main disadvantages. First, traditional methods have limited ability in representation learning as they  can only capture specific features such as self-similarity. Second, for most traditional denoisers, the calculation of derivatives for the denoising function would be a daunting task \cite{a33}, which may jeopardize the integration of DeepRED with neural networks.

\begin{figure}[htb]

  \centering
  \centerline{\includegraphics[width=8cm]{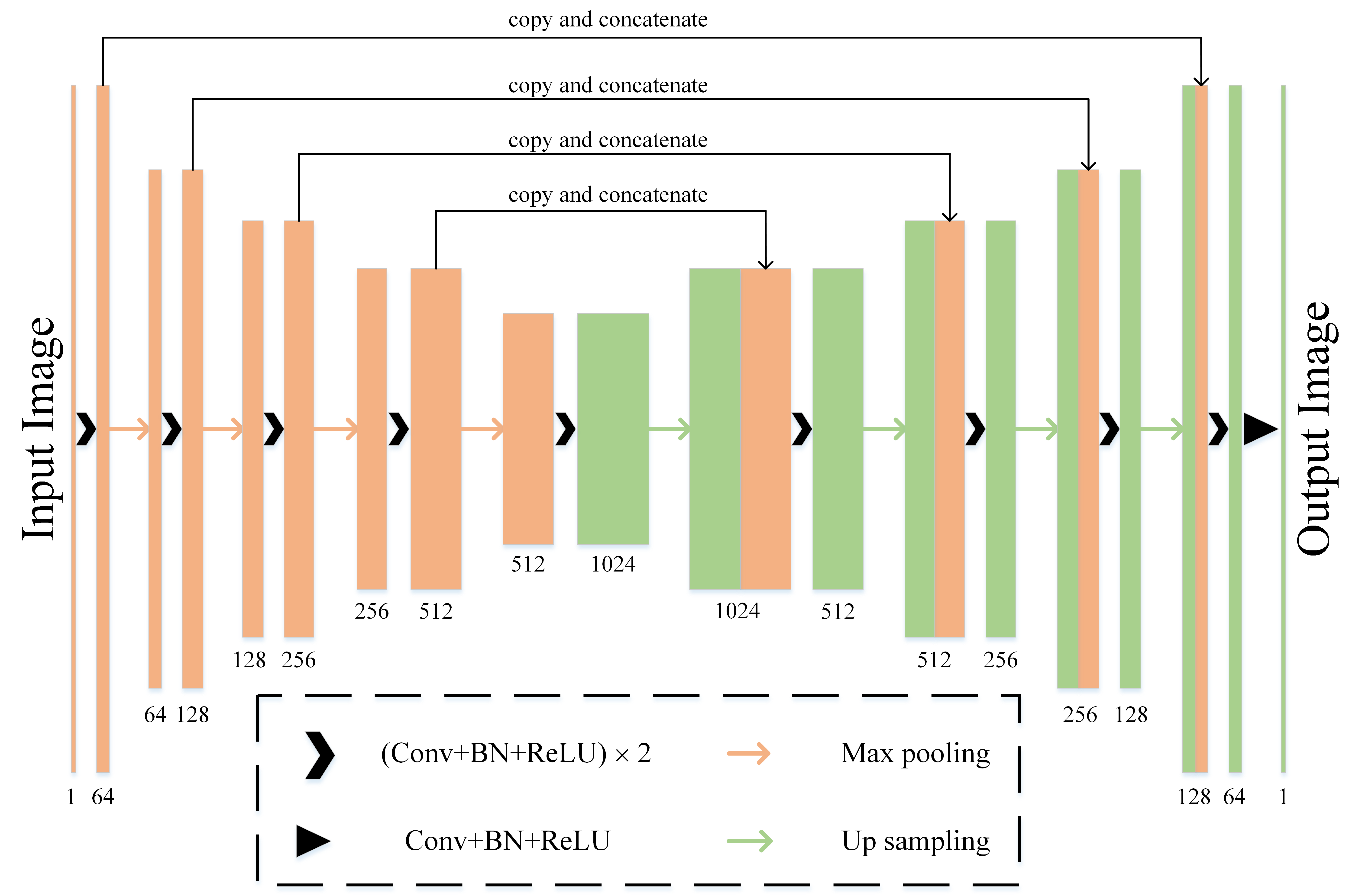}}
%  \vspace{2.0cm}
  \caption{Architecture of the generator network in our proposed model.}
  \label{fig:generator}
  \end{figure}

In this paper, we propose a PET reconstruction model combining the DeepRED framework with approximate Bayesian inference. Our model reconstructs PET images from a single sinogram without any anatomic prior, e.g., corresponding CT or MR images. In particular, the traditional denoiser commonly used in DeepRED is substituted with a learnable denoiser. As a result, the proposed model includes two CNNs: one is a generator that takes arbitrary input and outputs the reconstructed image, and the other plays the role of denoiser in DeepRED. Additionally, the denoiser learns to impose a task-based constraint on DIP and obtains adaptivity that does not exist in traditional denoisers. On the other hand, our denoiser is differentiable, which makes backpropagation possible to optimize the loss function. We also considered SGLD. Specifically, Gaussian noise is injected into the gradient updates, which derives a Markov chain Monte Carlo (MCMC) sampler from stochastic gradient descent (SGD), further alleviating overfitting from a Bayesian perspective. The main contributions of this work are summarized as follows: 1) an unsupervised end-to-end PET image reconstruction model under the framework of DeepRED, which does not need paired samples, is proposed. 2) A learnable denoiser embedded in DeepRED is used, which adds force to DIP adaptively and facilitates the calculation of the gradient. 3) DeepRED is combined with SGLD to further improve the regularization performance.

The remainder of this paper is organized as follows. In the next section, we introduce the preliminary knowledge of DIP and DeepRED. Sections III elaborate on the proposed method. In Section IV, brain data and whole-body data experiments are conducted to evaluate our method. Finally, a discussion and conclusions are given in Section V.

\section{Preliminaries}

\subsection{Deep Image Prior}
DIP is a state-of-the-art unsupervised framework that can restore corrupted images without training data. DIP suggests that the unknown image to be reconstructed $x$ can be represented using a carefully selected neural network $g$ as $x = g(\theta \mid z)$, where $\theta$ and $z$ denote the network parameter set and a random input, respectively. Consider an inverse problem $y = Px + n$, where $y$ is the measurement, $P$ represents the sampling matrix dependent on the specific task, and $n$ denotes additive noise, which is usually assumed to be Gaussian noise. Under the DIP framework, $x$ can be obtained first by optimizing:
\begin{equation}
    \begin{split}
      \hat{\theta} = arg\min _{\theta} \left\| P\cdot g(\theta \mid z) - y \right\|.
    \end{split}
  \end{equation}
Once $\hat{\theta}$ is obtained, $\hat{x}=g(\hat{\theta} \mid z)$ is obtained.

\begin{figure}[htb]

  \centering
  \centerline{\includegraphics[width=8cm]{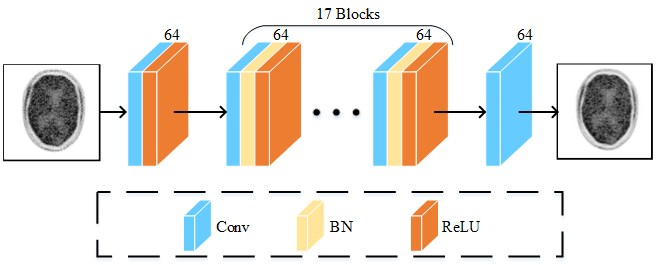}}
%  \vspace{2.0cm}
  \caption{Architecture of the denoiser network in our proposed model.}
  \label{fig:Denoiser}
  \end{figure}

\subsection{DeepRED}
DeepRED was proposed to mitigate overfitting by combining DIP and regularization by denoising. The objective function of DeepRED becomes:

\begin{equation}
    \begin{aligned}
    \min_{x,\theta} \quad & \left\| P\cdot g(\theta \mid z) - y \right\| +\frac{\lambda}{2}x^T(x-f(x)),\\
    \textrm{s.t.} \quad & x=g(\theta \mid z),\\
    \end{aligned}
    \end{equation}
where $f(\cdot)$ represents a denoiser such as NLM and BM3D, and $ \frac{\lambda}{2}x^T(x-f(x))$ is the so-called RED, which turns a denoiser into regularization. This framework relies on an image-adaptive Laplacian that poses cogent prior acting as a regularization term in various inverse problems \cite{a41}--\cite{a43}. This kind of regularization exhibits a better-managed optimization due to homogeneity and strong passivity on denoisers.

\begin{figure}[htb]

  \centering
  \centerline{\includegraphics[width=8cm]{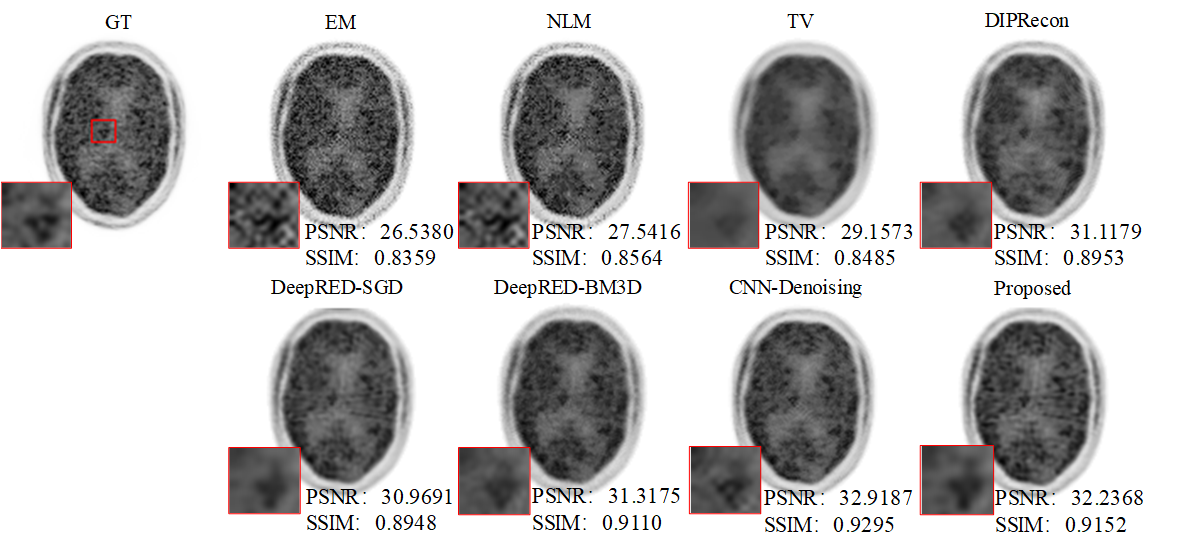}}
%  \vspace{2.0cm}
  \caption{Reconstructed axial brain PET image using different methods with a count level of $1 \times 10^8$ from one patient.}
  \label{fig:brain_axial}
  \end{figure}

  \begin{figure*}[htb]

    \centering
    \centerline{\includegraphics[width=17cm]{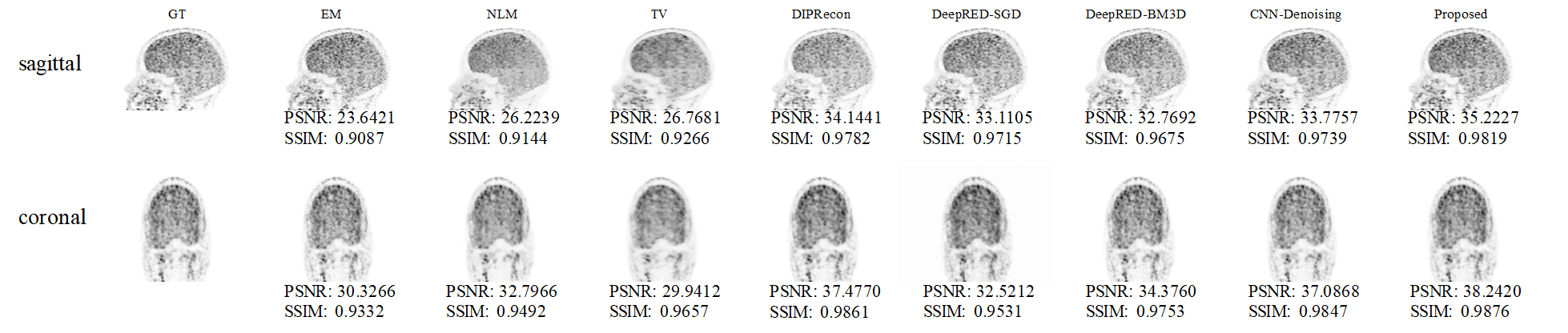}}
    %  \vspace{2.0cm}
    \caption{Reconstructed sagittal and coronal brain PET images using different methods with a count level of $1 \times 10^8$ from another patient.}
    \label{fig:brain_sagittal}
    \end{figure*}

\section{Method}

\subsection{PET Reconstruction Model}
The measured PET projection data $y\in \mathbb{R} ^{M}$ are known as sinograms, where each element $y_i$ is the number of photons accumulated in the $i$-th line of response (LOR). For $M$ lines of response and $N$ pixels in original image $x \in \mathbb{R}^{N}$, a typical PET imaging model is formulated as:
\begin{equation}
    \begin{split}
      \bar{y} = Px +r + s, 
    \end{split}
  \end{equation}
where $\bar{y} \in \mathbb{R} ^{M}$ is the expectation of $y$, $P \in \mathbb{R}^{M\times N}$ denotes the system matrix, $r \in \mathbb{R}^{M}$ and $s \in \mathbb{R}^{M}$ represent the random coincidences and scattered coincidences, respectively. Assuming that $y_i$ follows a distribution of $p(y_i \mid x)$, the log-likelihood for y can be written as
\begin{equation}
    \begin{split}
      L(x \mid y) = \sum_{i=1}^{M} log\,p(y_{i} \mid x).
    \end{split}
  \end{equation}
Under ideal conditions, $y_{i}$ is modeled as an independent Poisson variable. However, because of several acquisition circumstances, such as detector systems, electronic circumstances and various data preprocessing, the actual distribution often deviates from a Poisson distribution \cite{a39, a61}. Since several additive sources of error tend to form a Gaussian distribution, a Gaussian noise model is suitable for PET data. In fact, \cite{a40} shows that both Poisson and Gaussian models can be leveraged for PET reconstruction. In this article, to introduce the DeepRED framework, we adopt a Gaussian-based log-likelihood function as:
\begin{equation}
    \begin{split}
      p(y_i \mid x) = \frac{1}{\sqrt{2\pi}\sigma}exp\{ -\frac{(y_i - \bar{y_i})^2}{2\sigma ^2}\}. 
    \end{split}
  \end{equation}
Assuming that the variance of each LOR is equal, we have
\begin{equation}
    \begin{split}
      L(x \mid y) = - \left\|y-\bar{y}\right\|_2^2 + C,
    \end{split}
  \end{equation}
where $C$ is a constant that can be ignored.

\subsection{The Proposed Reconstruction Model}
In the proposed method, the reconstructed image $x$ is represented by:
\begin{equation}
    \begin{split}
      x = G(\theta_{G} \mid z),
      \label{dip}
    \end{split}
  \end{equation}
where $G$ denotes a generative network. Specifically, in this work, U-Net is employed as the backbone of $G$. $\theta_{G}$ represents the parameter set of $G$, and $z$ is the input of $G$. In contrast to \cite{a31}, in which other modality images, such as MRI or CT, are selected as $z$, we simply set $z=P^{T}y$ to keep our method more robust. By introducing a negative log-likelihood as a fidelity term into DeepRED, the objective function is obtained as
\begin{equation}
    \begin{split}
      \min_{x} {\left\|y-Px-r-s\right\|}^2_2 + \frac{\lambda}{2}x^T(x-f(x)). \label{obj}
    \end{split}
  \end{equation}
In the original DeepRED \cite{a33}, which uses NLM or BM3D as a denoiser, the derivative of $f(\cdot)$ with respect to $x$ is difficult to calculate. To circumvent this obstacle, ADMM is employed. However, ADMM introduces additional parameters to be tuned. Furthermore, traditional denoisers have demonstrated limited performance compared to recently proposed deep learning-based models. To further improve the denoising performance, we replace traditional denoisers with a DnCNN-like network \cite{a44} and denote it as $D$. Then, \eqref{obj} can be rewritten as

\begin{figure*}[htb]

  \centering
  \centerline{\includegraphics[width=17cm]{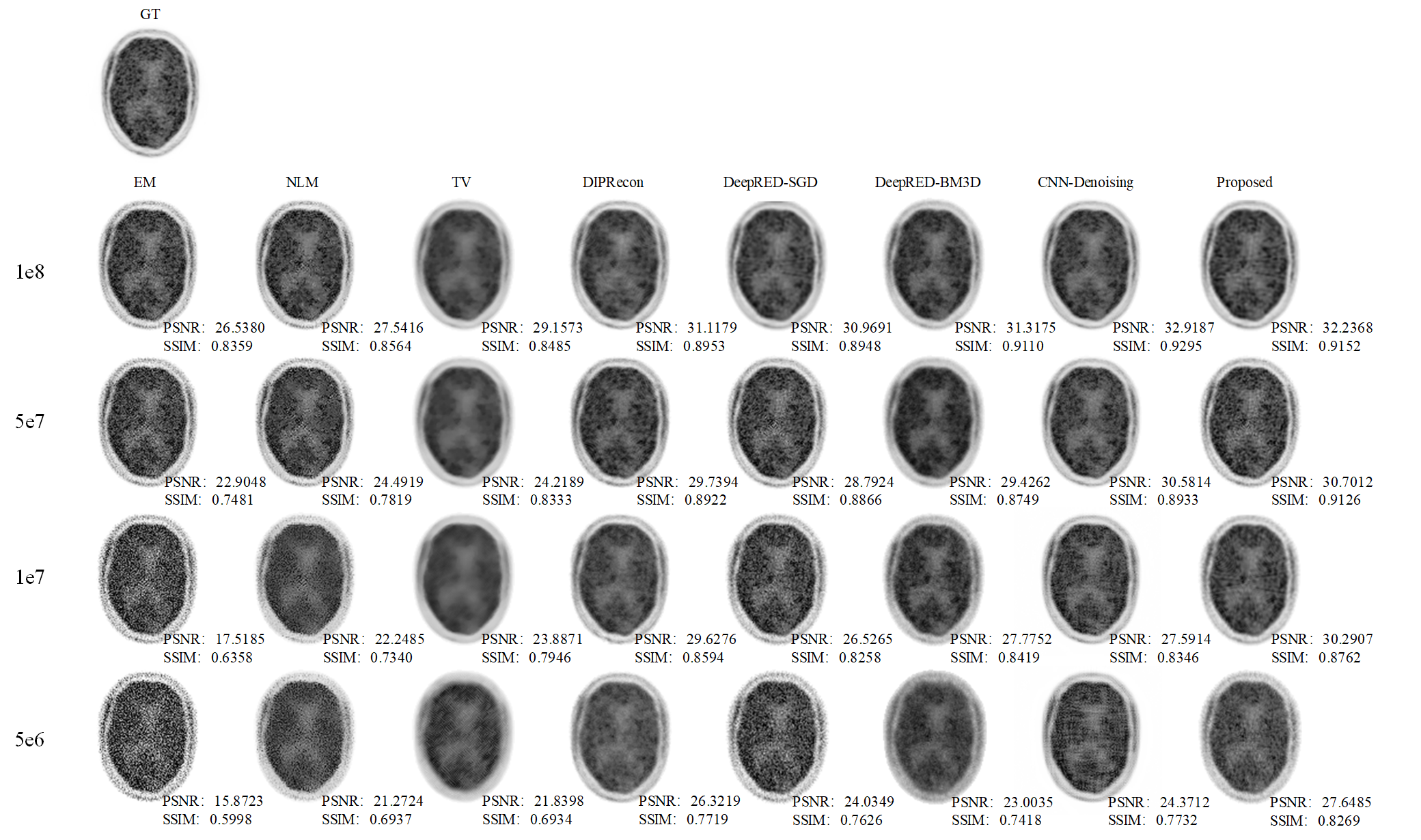}}
%  \vspace{2.0cm}
  \caption{Results with different methods with four count levels in the brain dataset study.}
  \label{fig:brain_level}
  \end{figure*}

\begin{equation}
    \begin{split}
      &\min_{\theta_{G}, \theta_{D}} {\left\|y-PG(\theta_G \mid z)-r-s\right\|}^2_2 \\
      + \frac{\lambda}{2} G(&\theta_G \mid z)^T(G(\theta_G \mid z)-D(\theta_D \mid G(\theta_G \mid z))), \label{obj2}
    \end{split}
  \end{equation}
where $\theta_{D}$ represents the parameter set of $D$. Then, we have two networks $D$ and $G$ in our model and will train them in an end-to-end manner.

\begin{figure}[htb]

  \centering
  \centerline{\includegraphics[width=8cm]{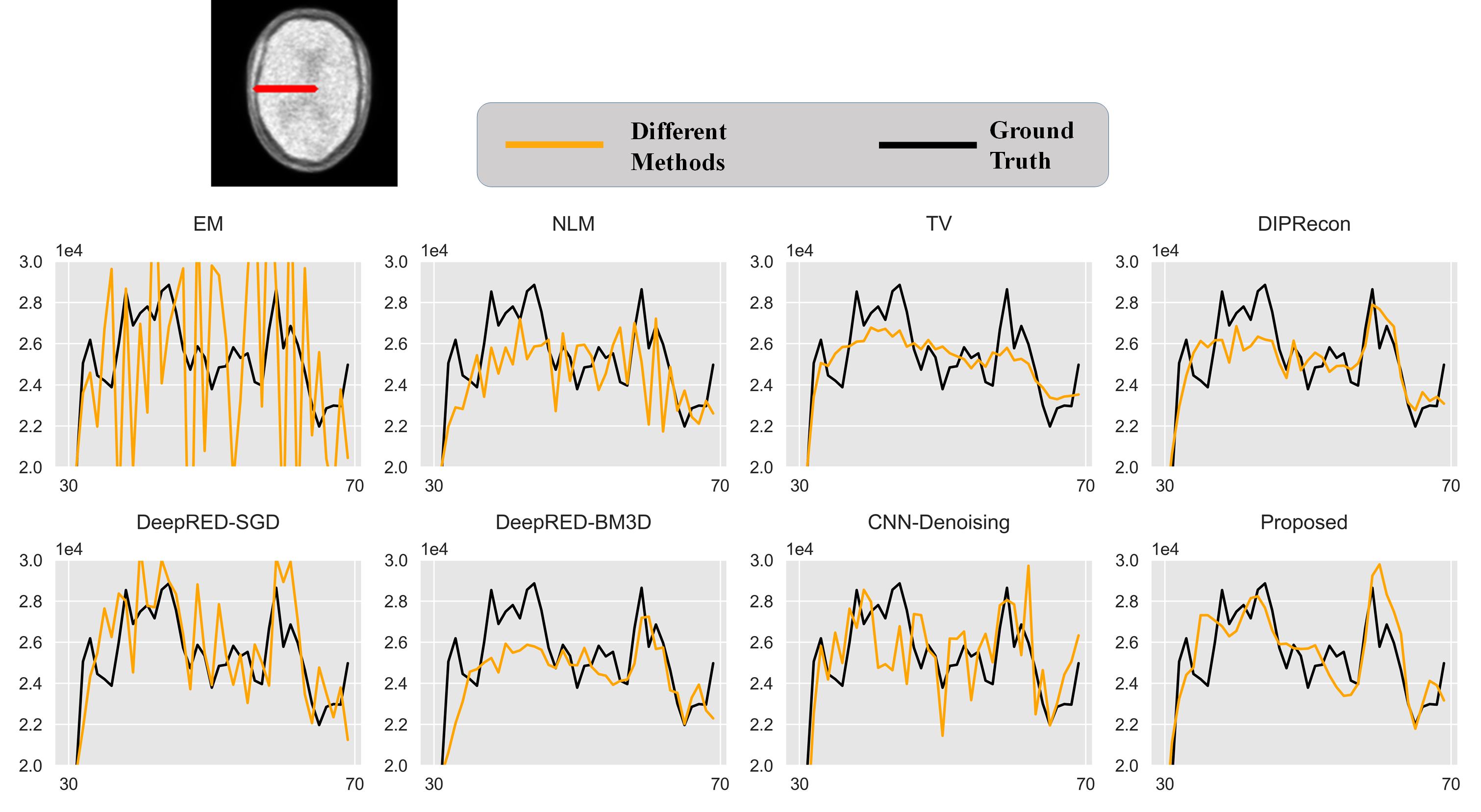}}
  %  \vspace{2.0cm}
  \caption{Comparison of horizontal line profiles in brain PET images. The horizontal axis represents the pixel index, while the vertical axis represents the pixel intensity.}
  \label{fig:brain_profile}
  \end{figure}

Although RED can mitigate overfitting to a certain degree, the optimization of \eqref{obj2} still leads to point estimation, which is apt to cause overfitting. To further relieve the overfitting and construct an overall regularization, we treat this problem from a Bayesian perspective. In Bayesian inference, a suitable prior distribution $p(\theta_G)$ is added to the parameter set $\theta_G$, and the final estimation $\hat{G}$ is formulated by marginalizing the posterior: $\hat{G}=\int G(\theta_{G} \mid z)p(\theta_{G} \mid y)d\theta_{G}$. A common solution to this posterior average is MCMC. However, MCMC converges more slowly than backpropagation while optimizing the network parameters. To handle this problem, SGLD was proposed for generating an MCMC sampler by injecting a proper amount of Gaussian noise in the updating step. SGLD formulates the updating equation as:
\begin{equation}
    \begin{split}
      \theta_{t+1} = \theta_{t} + \frac{\epsilon}{2} (-\nabla_{\theta_G}obj(\theta_G)+\nabla_{\theta_G}p(\theta_G)) + \eta_t,  \label{SGLD}
    \end{split}
  \end{equation}
where $\eta_{t} \sim \mathcal{N} (0, \epsilon)$ is the injected Gaussian noise, $\epsilon$ is the step size, and $obj(\cdot)$ denotes the objective function. $\nabla_{\theta_{G}}obj(\theta_{G})$ and $\nabla_{\theta_{G}}p(\theta_{G})$ represent the derivatives of $obj(\theta_{G})$ and $p(\theta_{G})$, respectively, with respect to $\theta_{G}$. The first term on the right-hand side of \eqref{SGLD} is the standard SGD updating, where $-\nabla_{\theta_{G}}obj(\theta_{G})$ can be considered a likelihood function, and $\nabla_{\theta_{G}}p(\theta_{G})$ can be treated as a prior, such as the L1 or L2 penalty. Theoretically, $\epsilon$ converges to zero. Since SGLD with diminishing step size converges weakly to a posterior distribution \cite{a49}, a constant rather than diminishing step size is commonly used in practice \cite{a50}--\cite{a52}. In this work, a constant step size specific to different images and count levels is adopted. The randomness comes from two sources in \eqref{SGLD}: one is the injected Gaussian noise, and the other is the noise in the stochastic gradient. In the early phase, the stochastic gradient takes control over the updating equation, leading SGLD to behave like SGD. In the later phase, the injected Gaussian noise dominates, and the algorithm steps into the Langevin dynamics phase, allowing estimation over the posterior phase.

Fig.\,\ref{fig:net} shows the overview of our proposed model. $y$ is the measured sinogram, which is the only data we have in our algorithm. $P^{T}y$ is used as the initial input to the generative network $G$. $x$ is the output of $G$. $\hat{y}$ is the corresponding sinogram generated from $x$. On the other hand, $x$ acts as the input to the denoiser $D$, and $D(x)$ denotes the output of $D$. '$\times$' denotes matrix multiplication. The final loss is composed of $x^{T}(x-D(x))$ and the Euclidean distance between $y$ and $\hat{y}$. The factor $\lambda$ controls the trade-off between two terms.

\begin{table}
  \caption{PSNR and SSIM Values of Various Methods of Brain Data Under a Count Level of $1 \times 10^8$.}
	\label{tab1}
\begin{center}
\resizebox{8.5cm}{!}{
\begin{tabular}{ccccccc}
\toprule
 & \multicolumn{2}{c}{Axial} &\multicolumn{2}{c}{Sagittal}&\multicolumn{2}{c}{Coronal} \\
Methods&PSNR&SSIM&PSNR&SSIM&PSNR&SSIM\\
\midrule
EM&26.5380&0.8359&23.6421&0.9087&30.3266&0.9332\\
NLM&27.5416&0.8564&26.2239&0.9144&32.7966&0.9492\\
TV&29.1573&0.8485&26.7681&0.9266&29.9412&0.9657\\
DIPRecon&31.1179&0.8953&34.1441&0.9782&37.4770&0.9861\\
DeepRED-SGD&30.9691&0.8948&33.1105&0.9715&32.5212&0.9531\\
DeepRED-BM3D&31.3175&0.9110&32.7692&0.9675&34.3760&0.9753\\
CNN-Denoising&\textbf{32.9187}&\textbf{0.9295}&33.7753&0.9739&37.0868&0.9847\\
Proposed&32.2368&0.9152&\textbf{35.2227}&\textbf{0.9819}&\textbf{38.2420}&\textbf{0.9876}\\
\bottomrule
\end{tabular}
}
\end{center}

\end{table}

\begin{table}
  \renewcommand\arraystretch{1.25}
  \caption{PSNR and SSIM Values of Various Methods of Brain Data Under Different Count Levels.}
  \label{tab2}
\begin{center}
\resizebox{8.5cm}{!}{
\begin{tabular}{ccccccccc}
\toprule
  & \multicolumn{2}{c}{$1 \times 10^8$} &\multicolumn{2}{c}{$5 \times 10^7$}&\multicolumn{2}{c}{$1 \times 10^7$}&\multicolumn{2}{c}{$5 \times 10^6$} \\
Methods&PSNR&SSIM&PSNR&SSIM&PSNR&SSIM&PSNR&SSIM\\
\midrule
EM&26.5380&0.8359&22.9048&0.7481&17.5185&0.6358&15.8723&0.5998\\
NLM&27.5416&0.8564&24.4919&0.7819&22.2485&0.7340&21.2724&0.6937\\
TV&29.1573&0.8485&24.2189&0.8333&23.8871&0.7946&21.8398&0.6934\\
DIPRecon&31.1179&0.8953&29.7394&0.8922&29.6276&0.8594&26.3219&0.7719\\
DeepRED-SGD&30.9691&0.8948&28.7924&0.8866&26.5265&0.8258&24.0349&0.7626\\
DeepRED-BM3D&31.3175&0.9110&29.4262&0.8749&27.7752&0.8419&23.0035&0.7418\\
CNN-Denoising&\textbf{32.9187}&\textbf{0.9295}&30.5814&0.8933&27.5914&0.8346&24.3712&0.7732\\
Proposed&32.2368&0.9152&\textbf{30.7012}&\textbf{0.9126}&\textbf{30.2907}&\textbf{0.8762}&\textbf{27.6485}&\textbf{0.8269}\\
\bottomrule
\end{tabular}
}
\end{center}

\end{table}

% \begin{figure*}[htb]

%   \centering
%   \centerline{\includegraphics[width=18cm]{brain_psnr.png}}
% %  \vspace{2.0cm}
%   \caption{Curves of PSNR and SSIM over count level in the brain dataset study.}
%   \label{fig:brain_panr}
%   %
%   \end{figure*}

\subsection{Network Architecture}

Fig.\,\ref{fig:generator} illustrates the architecture of the generative network $G$, which is derived from U-Net \cite{a21}. It contains four paired contracting and expansive blocks, and four skip connections are introduced between each pair. In the beginning, an image with a size of $128 \times 128$ is entered to $G$, followed by repetitive blocks that are uniformly composed of a $3 \times 3$ convolution layer, a batch normalization (BN) layer, a rectified linear unit (ReLU) layer and a $2 \times 2$ max pooling layer with stride 2. Each step in the expansive phase consists of a bilinear interpolated result concatenated with the corresponding feature map from the contracting step. Eventually, a $1 \times 1$ convolution is executed to compress the 64-channel feature map to a 1-channel output. The differences between our generative network and U-Net are as follows: 1) bilinear interpolation is used instead of deconvolution upsampling to reduce the checkerboard artifact; and 2) the padding operation is used after each $3 \times 3$ convolution to maintain the same size during propagation.
Fig.\,\ref{fig:Denoiser} displays the architecture of our denoising network $D$, which employs DnCNN as the backbone. The only difference between our model and DnCNN is that we exclude residual connections

\begin{figure*}[htb]

  \centering
  \centerline{\includegraphics[width=16cm]{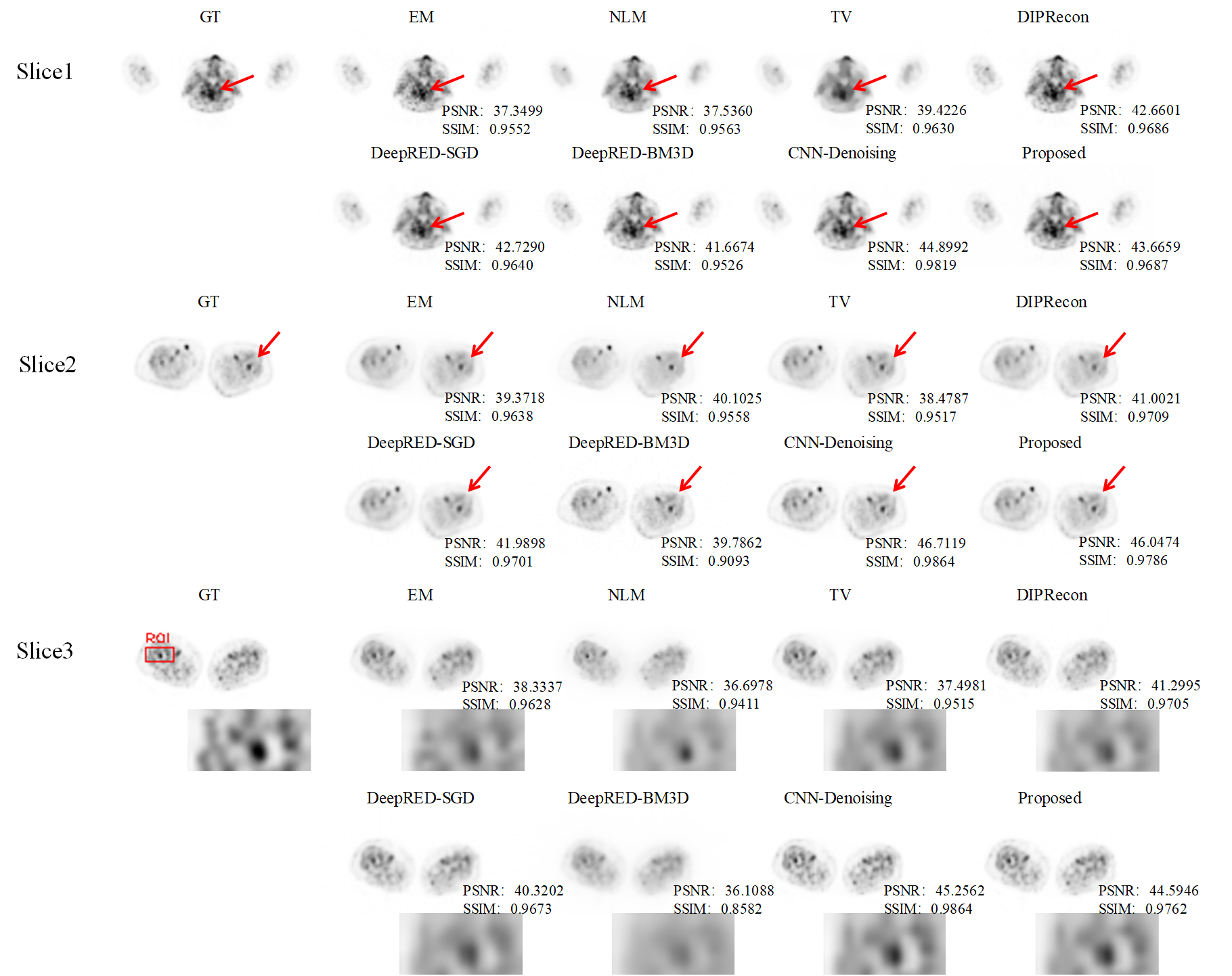}}
%  \vspace{2.0cm}
  \caption{Three axial slices of reconstructed whole-body PET images using different methods at a $1 \times 10^7$ count level.}
  \label{fig:whole_slice}
  \end{figure*}

\begin{figure}[htb]

  \centering
  \centerline{\includegraphics[width=9cm]{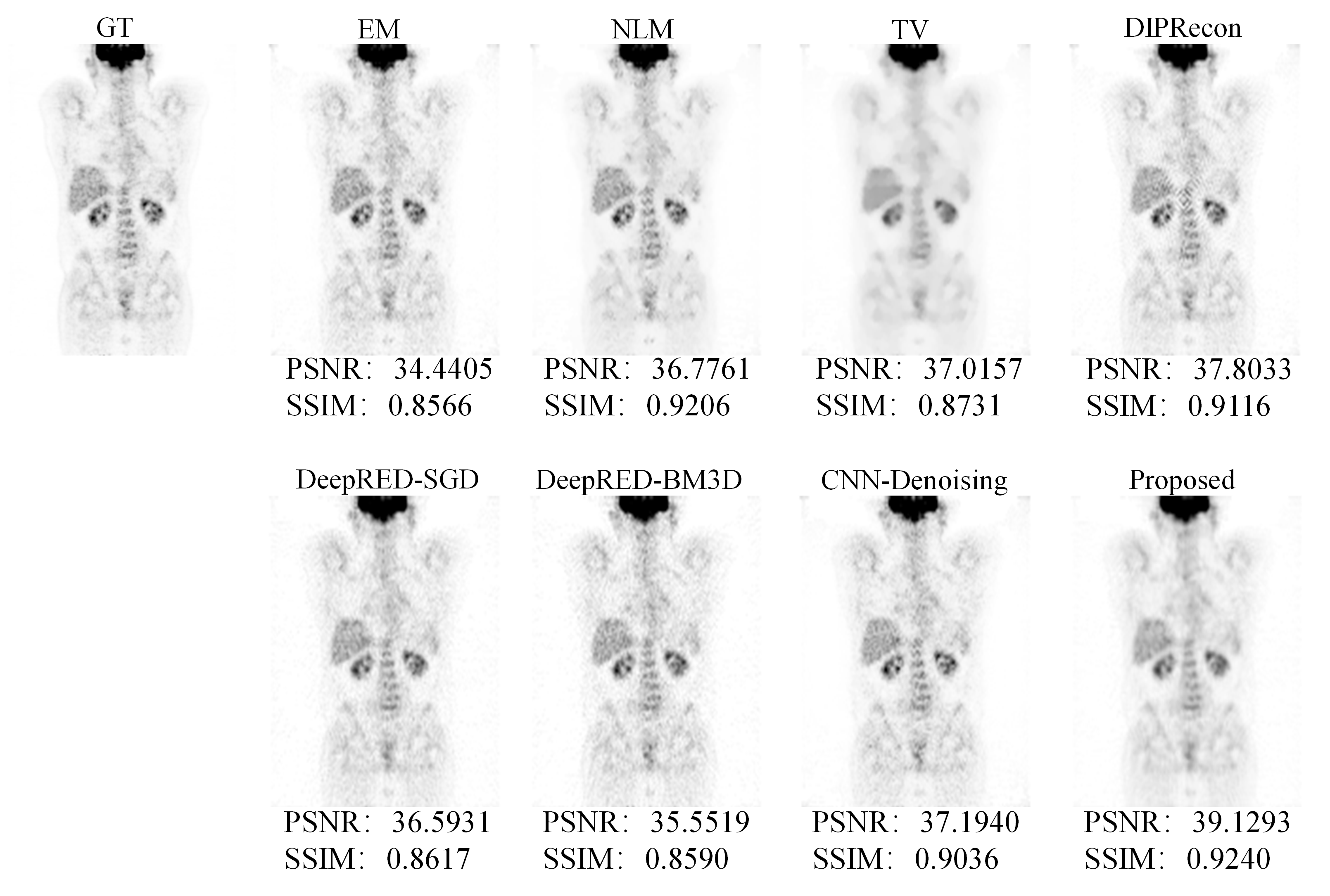}}
  %  \vspace{2.0cm}
  \caption{ Reconstructed axial and coronal whole-body PET images from another patient using different methods at a $1 \times 10^7$ count level.}
  \label{fig:whole_conoral}
  \end{figure}

\section{Experimental Setup and Results}

\begin{figure*}[htb]

  \centering
  \centerline{\includegraphics[width=16cm]{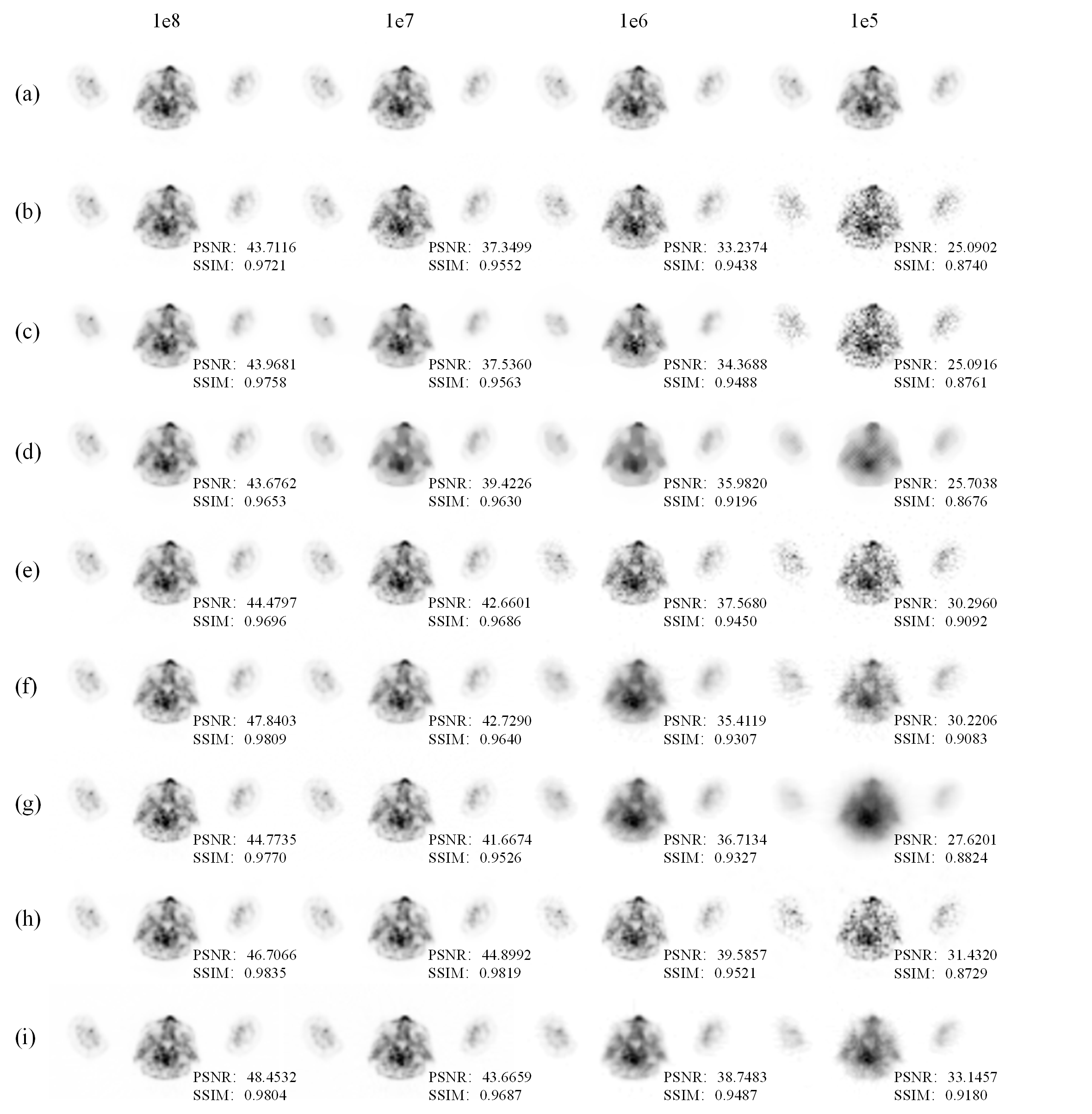}}
  %  \vspace{2.0cm}
  \caption{Results with different methods with four count levels in the whole-body dataset study. (a) Ground Truth, (b) EM, (c) NLM, (d) TV, (e) DIPRecon, (f) DeepRED-BM3D, (g) DeepRED-SGD, (h) CNN-Denoising, (i) Proposed.}
  \label{fig:whole_level}
  \end{figure*}

  \begin{figure}[htb]

    \centering
    \centerline{\includegraphics[width=8cm]{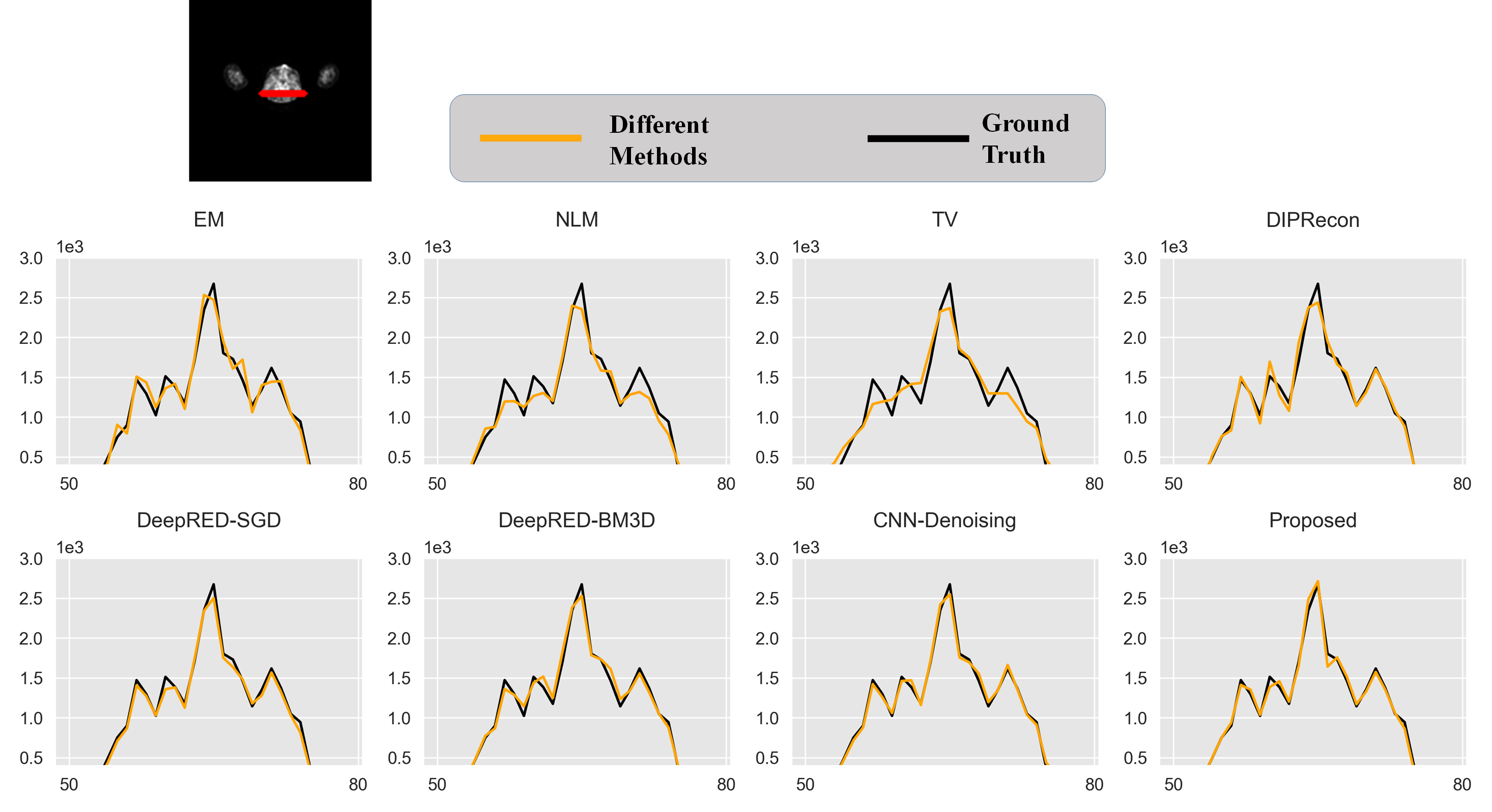}}
    %  \vspace{2.0cm}
    \caption{Comparison of horizontal line profiles in whole-body PET images. The horizontal axis represents the pixel index, while the vertical axis represents the pixel intensity.}
    \label{fig:whole_profile}
    \end{figure}

\subsection{Experimental Setup}
The PET brain images used in this part were downloaded from The Cancer Imaging Archive (TCIA) \cite{a54}. Two patients’ brain images were obtained using GE medical systems. One voxel size is $1.92 \times 1.92 \times 3.27 mm^{3}$, and the image size is $128 \times 128 \times 47$, the other voxel size is $4.06 \times 4.06 \times 5.00 mm^{3}$, and the image size is $128 \times 128 \times 145$. Each $FMISO$ PET imaging session consisted of acquiring a low-dose CT transmission scan used for attenuation correction and a PET emission scan at approximately two hours after injection of $^{18}FMISO$. The system matrix $P$ was modeled using Siddon’s refined method to calculate the ray path integral \cite{a47}. The noise-free sinogram was generated by forward projecting the PET image. Each sinogram has a matrix size of 180 (radial bins) $\times$ 180 (azimuthal angles). Then, uniform events were added to the noise-free data as background, which accounts for 10$\%$ of total true coincidences. Poisson noise was added by setting the total number of photons. In the brain dataset, four count levels were tested, including $1\times10^8$, $5\times10^7$, $1\times10^7$ and $5\times10^6$. To further evaluate the performance of our proposed method, a whole-body dataset also downloaded from TCIA was tested. The whole-body data of ten patients were acquired as approximately 300 slices with a size of $128 \times 128$. Before the examination, the patients underwent fasting for at least 6 hours, and the blood glucose of each patient was less than 11 mmol/L. Whole-body emission scans were acquired 60 minutes after the intravenous injection of 18F-FDG (4.44 MBq/kg, 0.12 mCi/kg), with patients in the supine position in the PET scanner. FDG doses and uptake times were 168.72-468.79 MBq (295.8±64.8 MBq) and 27-171 min (70.4±24.9 minutes), respectively. 18F-FDG with a radiochemical purity of 95$\%$ was provided. Patients were allowed to breathe normally during PET and CT acquisitions. Attenuation correction of PET images was performed using CT data with the hybrid segmentation method. We followed the same procedure as in a previous brain dataset study to generate low count sinograms. In the whole-body dataset, count levels of $1\times10^8$, $1\times10^7$, $1\times10^6$ and $1\times10^5$ were tested.

To quantify the performance of different methods, the peak signal-to-noise ratio (PSNR) and similarity structure index (SSIM) were employed. The proposed model was implemented by the PyTorch Python library \cite{a45} on a PC (Intel Xeon E5-2620 CPU and a GeForce GTX 1080Ti GPU). The network was trained using the Adam \cite{a46} optimizer, and the learning rate was set to 0.0001.
Six state-of-the-art methods were involved for comparison, including EM \cite{a5}, TV \cite{a9}, NLM \cite{a17}, DIPRecon \cite{a31}, DeepRED-BM3D \cite{a33} and CNN-Denoising \cite{a48}. EM is a commonly used reconstruction method deployed in most commercial scanners. TV is a classic regularization model usually employed in different inverse problems. NLM is an efficient image denoising method that has been applied in many imaging modalities. DIPRecon is based on a Poisson-based unsupervised iterative reconstruction whose regularization term is formulated as DIP. DeepRED is an efficient iteration framework equipped with a selectable denoiser, and in our experiment, the well-known BM3D is employed as the denoiser dubbed DeepRED-BM3D. CNN denoising is an end-to-end denoising network based on supervised learning. Moreover, to validate the effectiveness of SGLD, we also compare the proposed model with SGD, named DeepRED-SGD. Particularly, when we trained CNN-Denoising, we randomly selected 180 images for the brain and 300 images for the whole-body.

\begin{table}
  \tiny
  \renewcommand\arraystretch{0.35}
  \caption{PSNR and SSIM Values of Various Methods of Whole-body Data Under a Count Level of $1 \times 10^7$.}
	\label{tab3}
\begin{center}
\resizebox{8.5cm}{!}{
\begin{tabular}{ccccc}
\toprule[0.35pt]
 & \multicolumn{2}{c}{Axial} &\multicolumn{2}{c}{Coronal} \\
Methods&PSNR&SSIM&PSNR&SSIM\\
\midrule[0.35pt]
EM&38.3518&0.9606&34.4405&0.8566\\
NLM&38.1121&0.9510&36.7761&0.9206\\
TV&38.4664&0.9554&37.0157&0.8731\\
DIPRecon&41.6539&0.9700&37.8033&0.9116\\
DeepRED-SGD&41.6796&0.9671&36.5931&0.8617\\
DeepRED-BM3D&39.1874&0.9067&35.5519&0.8590\\
CNN-Denoising&\textbf{45.6224}&\textbf{0.9849}&37.1940&0.9036\\
Proposed&44.7693&0.9745&\textbf{39.1293}&\textbf{0.9240}\\
\bottomrule[0.35pt]
\end{tabular}
}
\end{center}

\end{table}

\begin{table}
  \renewcommand\arraystretch{1.25}
  \caption{PSNR and SSIM of Various Methods  of Whole-body Data Under Different Count Levels.}
  \label{tab4}
\begin{center}
\resizebox{8.5cm}{!}{
\begin{tabular}{ccccccccc}
\toprule
  & \multicolumn{2}{c}{$1 \times 10^8$} &\multicolumn{2}{c}{$1 \times 10^7$}&\multicolumn{2}{c}{$1 \times 10^6$}&\multicolumn{2}{c}{$1 \times 10^5$} \\
Methods&PSNR&SSIM&PSNR&SSIM&PSNR&SSIM&PSNR&SSIM\\
\midrule
EM&43.7116&0.9721&37.3499&0.9552&33.2374&0.9438&25.0902&0.8740\\
NLM&43.9681&0.9758&37.5360&0.9563&34.3688&0.9488&25.0916&0.8761\\
TV&43.6762&0.9653&39.4226&0.9630&35.9820&0.9196&25.7038&0.8676\\
DIPRecon&44.4797&0.9696&42.6601&0.9686&37.5680&0.9450&30.2960&0.9092\\
DeepRED-SGD&47.8403&0.9809&42.7290&0.9640&35.4119&0.9307&30.2206&0.9083\\
DeepRED-BM3D&44.7735&0.9770&41.6674&0.9526&36.7134&0.9327&27.6201&0.8824\\
CNN-Denoising&46.7066&\textbf{0.9835}&\textbf{44.8992}&\textbf{0.9819}&\textbf{39.5857}&\textbf{0.9521}&31.4320&0.8729\\
Proposed&\textbf{48.4532}&0.9804&43.6659&0.9687&38.7483&0.9487&\textbf{33.1475}&\textbf{0.9180}\\
\bottomrule
\end{tabular}
}
\end{center}

\end{table}

\subsection{Brain Data Results}
Figs.\,\ref{fig:brain_axial} and \,\ref{fig:brain_sagittal} show the reconstructed brain PET images from different patients using various methods with $1 \times 10^8$ count levels from axial, coronal and sagittal views, respectively. In general, methods based on CNNs garner better results than conventional methods that blur the image or produce severe noise. Note that images reconstructed by DeepRED-SGD gain a negligible advantage over DeepRED-BM3D, probably because we chose a relatively better result for DeepRED-BM3D since it produces oscillatory curves, while we chose the result for DeepRED-SGD when it converged to a stable solution. It can also be observed that our proposed method outperforms all the other traditional and unsupervised methods in terms of detail recovery and noise suppression. The results with different count levels are shown in Fig.\,\ref{fig:brain_level}. Consistent with the case of the $1 \times 10^8$ count level, CNN-based methods perform better than conventional methods, and our proposed method still achieves better results than other traditional and unsupervised methods. When the count level reaches $1 \times 10^7$, the performance of EM, DeepRED-SGD and DeepRED-BM3D drops significantly, while the other methods exhibit much less performance degradation. Our proposed method has visible improvement on DeepRED, which confirms the merit brought by SGLD. The quantitative results listed in TABLE \ref{tab1} and TABLE \ref{tab2} show coherent trends to the visual inspection that our method achieves the competitive scores in terms of both metrics to CNN-denoising better results than other methods. Pixel values along the red line in Fig.\,\ref{fig:brain_profile} of the ground truth versus the images reconstructed using different methods are plotted in Fig.\,\ref{fig:brain_profile}. The proposed method yields a profile closer to the ground truth than the other methods. It can be observed that our method has a more robust performance against the decline of the count level. Although CNN-denoising also has close performance to our method, it belongs to supervised learning and requires large amounts of perfectly paired samples, which are almost practically impossible.

  % \begin{figure*}[htb]

  %   \centering
  %   \centerline{\includegraphics[width=18cm]{whole_psnr.png}}
  %   %  \vspace{2.0cm}
  %   \caption{Curves of PSNR and SSIM over count level in the whole-body dataset study.}
  %   \label{fig:whole_psnr}
  %   %
  %   \end{figure*}
    
  \begin{figure*}[htb]
  
    \centering
    \centerline{\includegraphics[width=18cm]{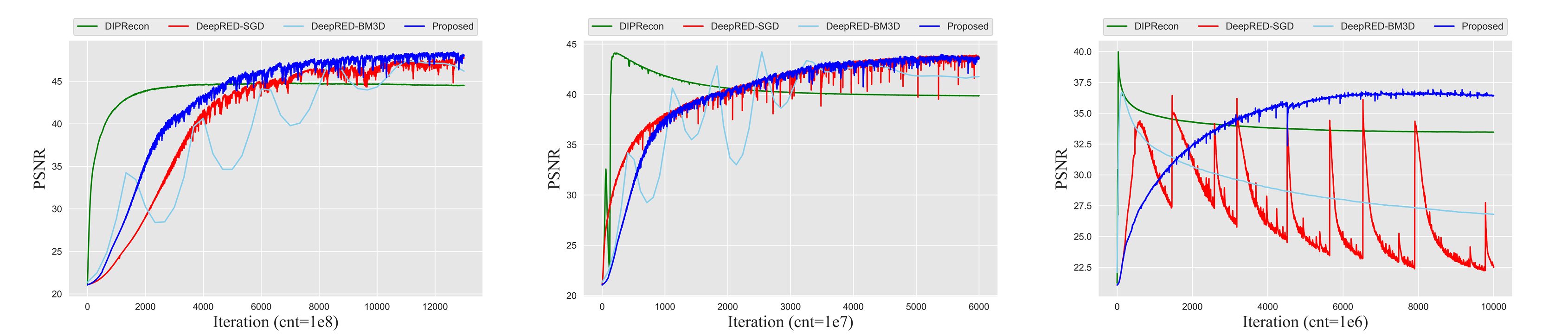}}
    %  \vspace{2.0cm}
    \caption{PSNR curves with the increase in iterations for DIP-based methods. Left: the count level of $1 \times 10^8$. Middle: the count level of $1 \times 10^7$. Right: the count level of $1 \times 10^6$.}
    \label{fig:convergence}
    \end{figure*}

\subsection{Whole-body Data Results}
Fig.\,\ref{fig:whole_slice} shows three typical whole-body slices from different body parts using different methods with count level of $1 \times 10^7$. For the first and second cases, red arrows indicate the regions that can be used to distinguish the performances of different methods. For the third case, a region of interest (ROI) marked by a red box is chosen to better visualize the improvement achieved by our method. Additionally, methods based on CNNs recover more details than conventional methods. Compared to DeepRED-BM3D, images reconstructed by DeepRED-SGD are closer to the ground truth, which agrees with the widely accepted fact that CNN-based denoisers generally bring better performance than some popular denoisers, such as BM3D. Notably, consisitent with the previous experiments on Brain Data, our method performs similar to CNN-Denoising, which belongs to supervised learning. Fig.\,\ref{fig:whole_conoral} shows a reconstructed whole-body image from a coronal view. Our method achieves the best visual effect, followed by DIPRecon, while DeepRED-SGD and DeepRED-BM3D generate much noise. Experiments on different count levels have been conducted, and the results are displayed in Fig.\,\ref{fig:whole_level}. As the count level decreases, the performance of each method drops significantly. When the count level reaches $1 \times 10^5$, DeepRED-BM3D can only reconstruct rough outlines without any details, and severe noise occurs in DIPRecon, while our method has a better trade-off between detail preservation and noise suppression than other methods. TABLE \ref{tab3} and TABLE \ref{tab4} show the means of PSNR and SSIM values in the whole-body dataset using different methods. Our method has better scores than all the other methods except CNN denoising. In addition, we select a line profile across the image, as shown in Fig.\,\ref{fig:whole_profile}, which further demonstrates that our proposed method obtains the best result. 
% Fig.\,\ref{fig:whole_psnr} shows the decline of PSNR and SSIM over count level.

\subsection{Convergence analysis}
The PSNR curves with the increase in iterations for each method are shown in Fig.\,\ref{fig:convergence}. Our method converges to a stable solution in all three count levels, but other methods produce oscillatory curves to different degrees. For a count level of $1 \times 10^8$, our proposed method achieves noticeable advantages over all the other methods. For a lower count level of $1 \times 10^7$, the scores of all the methods significantly decrease. DeepRED-SGD and our method yield similar performance and better performance than other methods. DIPRecon reaches the best result first but then declines, which indicates the occurrence of overfitting. When the count level is $1 \times 10^6$, DeepRED-SGD overfits while our method still converges, which shows the advantage brought by SGLD.

\section{Discussion and Conclusion}
In this work, we propose a PET reconstruction method directly from a sinogram under the DeepRED framework from a Bayesian perspective. Two well-known networks, U-Net and DnCNN, are selected as backbones of the generative network and denoiser in our framework, respectively. Notably, more state-of-the-art network components, such as octave convolution, attention modules and even transformers, may further improve the reconstruction performance. Since the main purpose of this work is to validate the effectiveness of our proposed framework, exploring the impacts of different network architectures will be our future work. In addition, only 2D cases are considered in this work as a pilot study. Introducing extra data from the spatial or temporal dimension will probably further improve the results, but this strategy will inevitably boost the computational cost. Since our method cannot be implemented in a patchwise manner (patch-based image denoising), 3D or 4D PET reconstruction will need much more video memory, which may be impractical for regular users, and light network architecture will be preferable for this situation.
Moreover, no anatomic prior is used in our method, which means our method is fully unsupervised. It was mentioned in \cite{a31} that taking MRI prior as complementary input could produce better results than random input. In future work, we will evaluate the performance of introducing anatomic images, such as CT or MRI images, as auxiliary priors.
The selection of parameters is also an important topic for our method. In the current approach, the parameters are manually optimized to fit the specific dataset or count level. How to develop an adaptive parameter selection strategy will also be a future topic of ours.

In conclusion, a PET reconstruction model that combines the DeepRED framework with approximate Bayesian inference is proposed in this paper. Our model restores PET images from a single sinogram in a fully unsupervised mode. A learnable denoiser, which has a differentiable form, is trained to substitute the conventional denoisers used in DeepRED. To further alleviate overfitting, Gaussian noise is injected into the gradient updates, which derives an MCMC sampler from SGD. Brain data and whole-body data experiments show that our proposed method produces better visual effects in detail preservation and noise suppression. Quantitative results also show that our proposed method outperforms several state-of-the-art methods in terms of PSNR and SSIM.


\begin{thebibliography}{00}

  \bibitem{a4} F. Natterer, \emph{The Mathematics of Computerized Tomography} (Classics in Applied Mathematics). Philadelphia, PA, USA: SIAM, 2001.

  \bibitem{a17} C. Chan \emph{et al.,} ``Postreconstruction nonlocal means filtering of whole-body PET with an anatomical prior,'' \emph{IEEE Trans. Med. Imag}., vol. 33, no. 3, pp. 636–650, Mar. 2014.
  
  \bibitem{a18}  S. Peltonen \emph{et al.,} ``PET sinogram denoising by block-matching and 3D filtering,'' in \emph{Proc. IEEE NSS/MIC}, Valencia, Spain, Oct. 2011, pp. 3125–3129.

  \bibitem{a68} J. Lin, A. F. Laine, and S. R. Bergmann, ``Improving pet-based physiological quantification through methods of wavelet denoising,” \emph{IEEE
  Trans. Biomed. Eng}., vol. 48, pp. 202–212, Feb. 2001.
  
  
  \bibitem{a5} L. A. Shepp and Y. Vardi, ``Maximum likelihood reconstruction for emission tomography,'' \emph{IEEE Trans. Med. Imag}., vol. 1, no. 2, pp. 113–122,
  Oct. 1982.
  
  \bibitem{a6} Z. Zhang \emph{et al.,} ``Investigation of optimization-based reconstruction with an image-total-variation constraint in PET,'' \emph{Phys. Med. Biol}., vol. 61, no. 16, p. 6055, 2016.
  
  \bibitem{a7} S. Ahn \emph{et al.,} ``Gap compensation during PET image reconstruction by constrained total variation minimization,'' \emph{Med. Phys}., vol. 39, no. 2, pp. 589-602, 2012.
  
  \bibitem{a8} A. Sawatzky \emph{et al.,} ``Accurate EM-TV algorithm in PET with low SNR,'' in \emph{Proc. Conf. Rec. IEEE Nucl. Sci. Symp}., Oct. 2008, pp. 5133–5137.
  
  \bibitem{a9} C. Wang \emph{et al.,} ``Low dose PET reconstruction with total variation regularization,'' in \emph{Proc. Eng. Med. Biol. Soc}., 2014, pp. 1917-1920.
  
  \bibitem{a10} G. Wang and J. Qi, ``Penalized likelihood PET image reconstruction using patch-based edge-preserving regularization,'' \emph{IEEE Trans. Med. Imag}, vol. 31, no. 12, pp. 2194–2204, 2012.
  
  \bibitem{a11} S. Ahn and R. Leahy, ``Analysis of resolution and noise properties of nonquadratically regularized image reconstruction methods for PET,'' \emph{IEEE Trans. Med. Imag}., vol. 27, pp. 413-424, Mar. 2008.
  
  \bibitem{a12} W. Zhang \emph{et al.,} ``Image reconstruction for positron emission tomography based on patch-based regularization and dictionary learning,'' \emph{Med.Phys}., vol. 46, no. 11, pp. 5014–5026, 2019.
  
  \bibitem{a13} G. Wang and J. Qi, ``PET image reconstruction using kernel method,'' \emph{IEEE Trans. Med. Imag}, vol. 34, no. 1, pp. 61–71, 2014.
  
  \bibitem{a14} S. Somayajula \emph{et al.,} ``PET image reconstruction using information theoretic anatomical priors,'' \emph{IEEE Trans. Med. Imag}., vol. 30, no. 3, pp. 537–49, Mar. 2011.
  
  \bibitem{a15} J. Cheng-Liao and J. Qi, ``PET image reconstruction with anatomical
  edge guided level set prior,'' \emph{Phys. Med. Biol}., vol. 56, no. 21, p. 6899,
  2011.
  
  \bibitem{a16} C. Chan \emph{et al.,} ``Regularized image reconstruction with an anatomically adaptive prior for positron emission tomography,'' \emph{Phys. Med. Biol}., vol. 54, no. 24, pp. 7379–400, 2009.
  

  \bibitem{a19} S. Ren \emph{et al.,} ``Faster R-CNN: Towards realtime object detection with region proposal networks,'' in \emph{Proc. Adv. Neural Inf. Process. Syst}., 2015, pp. 91–99.
  
  \bibitem{a20} K. He \emph{et al.,} ``Deep residual learning for
  image recognition,'' in \emph{Proc. IEEE Conf. Comput. Vis. Pattern Recognit}.,
  2016, pp. 770–778.
  
  \bibitem{a21} O. Ronneberger, P. Fischer, and T. Brox, ``U-Net: Convolutional networks for biomedical image segmentation,'' in \emph{Proc. Int. Conf. Med. Image Comput. Comput.-Assist. Intervent}., 2015, pp. 234–241.
  
  \bibitem{a22} H. Chen \emph{et al.,} ``Low-dose CT with a residual encoder-decoder convolutional neural network,'' \emph{IEEE Trans. Med. Imag}., vol. 36, no. 12,
  pp. 2524–2535, Dec. 2017.
  
  \bibitem{a23} Q. Yang \emph{et al.,} ``Low-dose CT image denoising using a generative
  adversarial network with wasserstein distance and perceptual loss,'' \emph{IEEE
  Trans. Med. Imag}., vol. 37, no. 6, pp. 1348–1357, Jun. 2018.

  \bibitem{a55} M. Ran \emph{et al.,} ``Denoising of 3D magnetic resonance images using a residual encoder--decoder Wasserstein generative adversarial network,'' \emph{Med. Image Anal}., vol. 55, pp. 165-180, 2019.
  
  \bibitem{a56} M. Ran \emph{et al.,} ``Md-recon-net: A parallel dual-domain convolutional neural network for compressed sensing MRI,'' \emph{IEEE Trans. Radiat. Plasma Med. Sci}, vol. 5, no. 1, pp. 120-135, 2020.
  
  \bibitem{a57} W. Xia \emph{et al.,} ``CT Reconstruction with PDF: Parameter-Dependent Framework for Data from Multiple Geometries and Dose Levels,'' \emph{IEEE Trans. Med. Imag}., 2021.
  
  \bibitem{a58} W. Xia \emph{et al.,} ``MAGIC: Manifold and Graph Integrative Convolutional Network for Low-Dose CT Reconstruction,'' \emph{IEEE Trans. Med. Imag}., 2021.
  
  \bibitem{a24} I. H{\"a}ggstr{\"o}m \emph{et al.,} ``DeepPET: A deep encoder–decoder network for directly solving the PET image reconstruction inverse problem,'' \emph{Med. Image Anal}., vol. 54, pp. 253–262, May. 2019.
  
  \bibitem{a25} Z. Hu \emph{et al.,} ``DPIR-Net: direct PET image reconstruction based on the Wasserstein generative adversarial network,'' \emph{IEEE Trans. Radiat. Plasma Med. Sci}., vol. 5, no. 1, pp. 35-43, 2020.
  
  \bibitem{a26}  K. Gong \emph{et al.,} ``PET Image Denoising Using a Deep Neural Network Through Fine Tuning,'' \emph{IEEE Trans. Radiat. Plasma Med. Sci}, vol. 3, no. 2, pp. 153-161, Mar. 2019.
  
  \bibitem{a27} C. C. Liu and J. Qi, ``Higher SNR PET image prediction using a deep
  learning model and MRI image,'' \emph{Phys. Med. Biol}., vol. 64, no. 11,
  May 2019, Art. no. 115004.

  \bibitem{a62} A. Mehranian \emph{et al.,} ``Image enhancement of whole-body oncology [18 F]-FDG PET scans using deep neural networks to reduce noise,'' \emph{Eur. J. Nucl. Med. Mol. Imaging}., 2021.

  \bibitem{a64} L. Zhou \emph{et al.,} ``Supervised learning with cyclegan for low-dose FDG PET image denoising,'' \emph{Med. Image Anal}., vol. 65, p. 101770, 2020.
  
  \bibitem{a28} K. Gong \emph{et al.,} ``Iterative PET Image Reconstruction Using Convolutional Neural Network Representation," \emph{IEEE Trans. Med. Imag}., vol. 38, no. 3, pp. 675-685, Mar. 2019.
  
  \bibitem{a29} A. Mehranian and A. J. Reader ``Model-based deep learning PET image reconstruction using forward-backward splitting expectation maximisation'' \emph{IEEE Trans. Radiat. Plasma Med. Sci}, vol. 5, no. 1, pp. 54-64, 2020.

  \bibitem{a65} I. Y. Chun and J. A. Fessler, ``Deep BCD-net using identical encodingdecoding CNN structures for iterative image recovery,” in \emph{Proc. Image, Video, and Multidim. Signal Process. Workshop}., Zagori, Greece, Apr. 2018, pp. 1–5.
  
  \bibitem{a66} H.Lim \emph{et al.,} ``Improved low-count quantitative PET reconstruction with an iterative neural network,'' \emph{IEEE Trans. Med. Imag}., vol. 39, no. 11, pp. 3512-3522, 2020.
  
  
  \bibitem{a30} D. Ulyanov \emph{et al.,} ``Deep image prior,'' in \emph{Proc. IEEE Conf. Comput. Vis. Pattern Recognit}., 2018, pp. 9446-9454.
  
  \bibitem{a31} K. Gong \emph{et al.,} ``PET image reconstruction using deep image prior,'' \emph{IEEE Trans. Med. Imag}., vol. 38, no. 7, pp. 1655-1665, Jul. 2019.
  
  \bibitem{a32} T. Yokota \emph{et al.,} ``Dynamic PET image reconstruction using nonnegative matrix factorization incorporated with deep image prior,'' in \emph{Proc. IEEE/CVF Int. Conf. Comput. Vis}., Oct. 2019, pp. 3126-3135.

  \bibitem{a67} J. Cui \emph{et al.,} ``PET image denoising using unsupervised deep learning'' \emph{Eur. J. Nucl. Med. Mol. Imaging}., vol. 46, no. 13, pp. 2780-2789, 2019.
  
  \bibitem{a33}  G. Mataev, P. Milanfar, and M. Elad, ``DeepRED: Deep image prior
  powered by red,'' in \emph{Proc. IEEE/CVF Int. Conf. Comput. Vis. Workshops}.,
  Oct. 2019, pp. 1–10.
  
  \bibitem{a34} Y. Romano, M. Elad, and P. Milanfar, ``The little engine that could:
  Regularization by denoising (RED),'' \emph{SIAM J. Imag. Sci}., vol. 10, no. 4,
  pp. 1804–1844, Jan. 2017.
  
  \bibitem{a35} Z. Cheng \emph{et al.,} ``A Bayesian perspective on the deep image prior,'' in \emph{Proc. IEEE Conf. Comput. Vis. Pattern Recognit}., Jun. 2019, pp. 5443-5451
  
  \bibitem{a36} M. Welling and Y. W. Teh, ``Bayesian learning via stochastic gradient langevin dynamics,'' in \emph{Proc. Int. Conf. Mach. Learn}., 2011, pp. 681–688.
  
  \bibitem{a37} H. Sun \emph{et al.,} ``Dynamic PET Image Denoising Using Deep Image Prior Combined With Regularization by Denoising,'' \emph{IEEE Access}., vol. 9, pp. 52378-52392, 2021.
  
  \bibitem{a38} H. Carrillo \emph{et al.,}  ``Low-count PET image reconstruction with Bayesian inference over a Deep Prior,'' in \emph{Med. Imag. 2021: Imag. Proc.}, vol. 11596, pp.115960V, 2021.

  \bibitem{a41} S. M. Haque, G. Pai, and V. M. Govindu, ``Symmetric smoothing filters from global consistency constraints,'' \emph{IEEE Trans. Image Process}.,
  vol. 24, no. 5, pp. 1536–1548, May 2014.
  
  \bibitem{a42} A. Kheradmand and P. Milanfar, ``A General Framework for Regularized, Similarity-Based Image Restoration,'' \emph{IEEE Trans. Image Process}., Vol. 23, No. 12, pp. 5136-5151, Dec. 2014.
  
  \bibitem{a43} Y. Romano and M. Elad, ``Boosting of Image Denoising Algorithms,'' \emph{SIAM J. Imag. Sci}., vol. 8, no. 2, pp. 1187-1219, Jun. 2015.
  
  \bibitem{a39} D. L. Bailey \emph{et al.,} \emph{Positron emission tomography}. London: Springer, 2005, p. 71.

  \bibitem{a61} Y. C. Cavalcanti \emph{et al.,} ``Factor analysis of dynamic PET images: beyond Gaussian noise,'' \emph{IEEE Trans. Med. Imag}., vol. 38, no. 9, pp. 2231-2241, 2019.
  
  \bibitem{a40} X. Yu \emph{et al.,} ``Low dose PET image reconstruction with total
  variation using alternating direction method,'' \emph{PloS one}., vol. 11, no. 12, p. e0166871, 2016.
  

  
  \bibitem{a44} K. Zhang \emph{et al.,} ``Beyond a Gaussian Denoiser: Residual learning of deep CNN for image denoising,'' \emph{IEEE Trans. Image Process}., vol. 26, no. 7, pp. 3142–3155,
  Jul. 2017.

  \bibitem{a49} N. Brosse, A. Durmus and E. Moulines, ``The promises and pitfalls of Stochastic Gradient Langevin Dynamics'' in \emph{Proc. Adv. Neural Inf. Process. Syst}., 2018, pp. 8268–8278.
  
  \bibitem{a50} S. Ahn, A. K. Balan, and M. Welling. ``Bayesian posterior sampling via stochastic gradient Fisher scoring,'' in \emph{Proc. Int. Conf. Mach. Learn}., 2012.
  
  \bibitem{a51} T. Chen, E. Fox, and C. Guestrin. ``Stochastic gradient hamiltonian Monte Carlo,'' in \emph{Proc. Int. Conf. Mach. Learn}., 2014, pp. 1683–1691.
  
  \bibitem{a52} C. Li \emph{et al.,} ``Preconditioned stochastic
  gradient Langevin dynamics for deep neural networks,'' in \emph{Proc. 30th
  AAAI Conf. Artif. Intell}., 2016, pp. 1788–1794.
  
  % \bibitem{a53} S. J. Vollmer, K. C. Zygalakis, and Y. W. Teh. ``Exploration of the (non-) asymptotic bias and variance of stochastic gradient langevin dynamics,'' \emph{J. Mach. Learn. Res}., vol. 17, no. 1, pp. 5504–5548, 2016.

  \bibitem{a54} K. Clark \emph{et al.,} ``The cancer imaging archive (TCIA): Maintaining and
  operating a public information repository,'' \emph{J. Digit. Imag}., vol. 26, no. 6,
  pp. 1045–1057, 2013.

  \bibitem{a47} G. Han, Z. Liang, and J. You, ``A fast ray-tracing technique for TCT and
  ECT studies,” in \emph{Proc. IEEE Nucl. Sci. Symp}., Seattle, WA, USA, 1999,
  vol. 3, pp. 1515–1518.
  
  
  \bibitem{a45} A. Paszke \emph{et al.,} ``Pytorch: An imperative style, high-performance deep learning library'' in \emph{Proc. Adv. Neural Inf. Process. Syst}., 2019, pp. 8026-8037.
  
  \bibitem{a46} D. P. Kingma and J. Ba, ``Adam: A method for stochastic optimization,” 2014, \emph{arXiv:1412.6980}. [Online]. Available: \underline{http://arxiv.org/abs/1412.6980}.
  

  
  \bibitem{a48} K. Gong \emph{et al.,} ``PET image denoising using a deep neural network through fine tuning,'' \emph{IEEE Trans. Radiat. Plasma Med. Sci}., vol. 3, no. 2, pp. 153–161, Mar. 2019.
  

  

  
  

  

  
  % \bibitem{a63} Y. Lei \emph{et al.,} ``Whole-body PET estimation from low count statistics using cycle-consistent generative adversarial networks,'' \emph{Phys. Med. Biol}., vol. 64, no. 21, p. 215017, Nov. 2019.
  

  

  

  

  
\end{thebibliography}
\end{document}